\documentclass[aps,prl,twocolumn,psfig,showpacs]{revtex4}
\usepackage{amsfonts}
\usepackage{amsmath}
\usepackage{graphicx}
\usepackage{bm}
\usepackage{color}
\usepackage{amssymb}
\usepackage{ulem}
\usepackage{times}
\usepackage{dcolumn}
\usepackage{cases}
\usepackage{txfonts}

\begin{document}

\title{Modified Tucker Decomposition for Tensor Network and Fast Linearized Tensor
Renormalization Group Algorithm for Two-Dimensional Quantum Spin
Lattice Systems}
\author{Shi-Ju Ran, Wei Li, and Gang Su}
\email[Corresponding author. ]{Email: gsu@gucas.ac.cn}
\affiliation{Theoretical Condensed Matter Physics and Computational
Materials Physics Laboratory, College of Physical Sciences, Graduate
University of Chinese Academy of Sciences, P. O. Box 4588, Beijing
100049, China}

\begin{abstract}
We propose a novel algorithm with a modified Tucker decomposition
for tensor network that allows for efficiently and precisely
calculating the ground state and thermodynamic properties of
two-dimensional (2D) quantum spin lattice systems, and is coined as
the fast linearized tensor renormalization group (fLTRG). Its
amazing efficiency and precision are examined by studying the
spin-$1/2$ anisotropic Heisenberg antiferromagnet on a honeycomb
lattice, and the results are found to be fairly in agreement with
the quantum Monte Carlo calculations. It is also successfully
applied to tackle a quasi-2D spin-$1/2$ frustrated bilayer honeycomb
Heisenberg model, where a quantum phase transition from an ordered
antiferromagnetic state to a gapless quantum spin liquid phase is
found. The thermodynamic behaviors of this frustrated spin system
are also explored. The present fLTRG algorithm could be readily
extended to other quantum lattice systems.
\end{abstract}

\maketitle

Efficient and accurate numerical methods are very crucial to tackle
the strongly correlated quantum lattice systems. Although some
analytical techniques and numerical methods have been proposed in
the past decades, a large class of intriguing correlated electron
and spin models are still intractable owing to the complexity of
quantum many-body systems. Several numerical renormalization group
(RG) approaches were thus developed, where the density matrix
renormalization group \cite{DMRG} and its finite temperature
variant---the transfer matrix renormalization group \cite{TMRG} have
achieved a great success for one-dimensional (1D) systems. Very
recently, generalizing the RG-based algorithms to two-dimensional
(2D) quantum lattice systems has been remarkably advancing. A few
numerical approaches, for instance, the projected entangle pair
state (PEPS) \cite{PEPS}, the tree tensor network (TN)
 \cite{TTN}, the multiscale entanglement renormalization
ansatz state \cite{MERA}, the infinite PEPS \cite{projection,
projection2}, the tensor renormalization group \cite{TRG1, TRG2,
TRG3}, and so on, were proposed, some of which already gained
interesting applications (e.g. Refs. [\onlinecite{Li, Chen}]). It is
noted that most of these algorithms are effective for the ground
state properties, but they are still difficultly applied to study
the thermodynamics of 2D quantum lattice models.

By incorporating the infinite time-evolving block decimation
technique \cite{iTEBD}, we developed a linearized TRG (LTRG)
algorithm \cite{LTRG} that renders a convenient way to investigate
the thermodynamic properties of low-dimensional quantum spin lattice
systems. Although the LTRG method is quite efficient and accurate
for 1D quantum systems, its cost is relatively high and the
performance near a critical point needs careful improvements for 2D
quantum systems. Within the framework of LTRG, when the density
operator is represented by a TN through Trotter-Suzuki decomposition
\cite{Trotter}, the truncation is needed to prevent from the
divergence of dimension of Hilbert space during the imaginary time
evolution, which will unavoidably bring errors that become worse in
2D quantum systems. To solve this issue, we note that the Tucker
decomposition (TD) \cite{Tucker,TensorD} is a nice way to obtain the
best lower dimensional approximation of a single tensor, and has
wide applications in areas of data compression, image processing, etc.
\cite{TensorD} The algorithms for the TD like the higher-order
singular value decomposition \cite{HOSVD,HOSVD1} and higher-order
orthogonal iteration (HOOI) \cite{HOOI} were suggested.

In this work, by extending the HOOI scheme to a TN instead of a
single tensor for an optimal truncation, we propose a novel
algorithm that allows us to efficiently and accurately simulate not
only the ground state but also thermodynamic properties of 2D
quantum spin lattice systems in the thermodynamic limit, which is
dubbed as the fast LTRG (fLTRG). We find that the computational cost
of fLTRG is insensitive to the coordination number without
losing the accuracy, which allows for a higher bond dimension cutoff
$D_{c}$ when it is applied to 2D and quasi-2D quantum systems. The
cost of the fLTRG algorithm is $\sim O(zD_{c}^{3})$, while the LTRG
is $\sim O(D_{c}^{3z-3})$, where $z$ is the coordination number. The
reliability, efficiency and accuracy of the fLTRG algorithm are
examined by studying a spin-1/2 anisotropic Heisenberg
antiferromagnet on the honeycomb lattice, whose energy per site,
staggered magnetization and specific heat are efficiently and
accurately obtained by the fLTRG, and the results are in good agreement with
quantum Monte Carlo (QMC) results. To show its powerful performance
and flexible scalability, we applied the fLTRG algorithm to a
spin-1/2 frustrated bilayer honeycomb Heisenberg model to which the
QMC is not directly accessible, and disclosed a quantum phase
transition (QPT) from an ordered antiferromagnetic phase to a
gapless quantum spin liquid (QSL) in the ground state. The
thermodynamic properties of this frustrated spin system are also
calculated. Our results manifest that the fLTRG would be very
promising to tackle the intractable correlated quantum many-body
systems in two and higher dimensions. In what follows, we shall
describe the basic procedure of the fLTRG algorithm with a quantum
spin system as an example on a honeycomb lattice.

\begin{figure}[tbp]
\includegraphics[angle=0,width=1\linewidth]{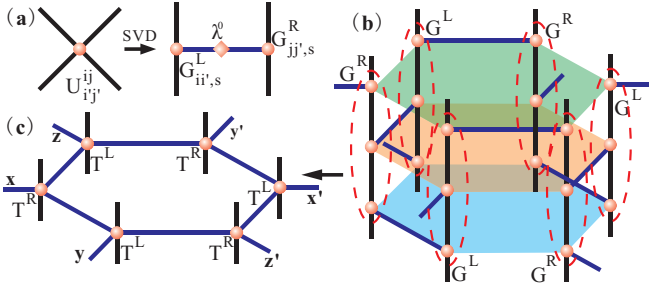}
\caption{(Color online) (a) The local evolution operator
$U^{ij}_{i'j'}$ is decomposed via an SVD into two gates,
$G^{L}_{ij,s}$ and $G^{R}_{i'j',s}$, each of which has two physical
bonds ($i,i'$ and $j,j'$, black) and one geometrical bond ($s$,
blue); (b) Contract the shared physical bonds among $G^L$ and $G^R$
to get tensors $T^L$ and $T^R$; (c) A TPDO with inverse temperature
$\tau$. Note that the singular value vectors $\lambda^{I,II,III}$ on
each geometrical bond is not indicated in (b) and (c) for concise.}
\label{fig-TensorNet}
\end{figure}

\begin{figure}[tbp]
\includegraphics[angle=0,width=1\linewidth]{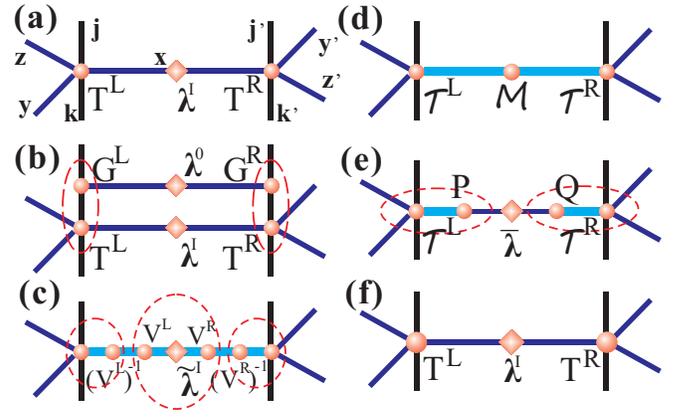}
\caption{(Color online) (a) A TPDO on bond $x$; (b) acting $G^{L}$
and $G^{R}$ on (a); (c) contracting the non-orthogonal
transformation matrices $V^{L}$ and $V^{R}$; (d) obtaining new
tensors $\mathcal{T}^{L}$, $\mathcal{T}^{R}$ and matrix ${\cal M}$;
(e) decomposing ${\cal M}$ by employing the SVD and contracting the
top $D_{c}$ left and right singular vectors $\mathcal{T}^{L}$ and
$\mathcal{T}^{R}$, respectively, to obtain the truncated $T^{L}$ and
$T^{R}$; and (f) keeping the $D_{c}$ largest singular values as new
$\lambda^{I}$.} \label{fig-HOOI}
\end{figure}

\textit{Initialization}.--- Suppose that the Hamiltonian of the
system can be written as $H=\sum_{i,j}\hat{H}_{ij}$, where
$\hat{H}_{ij}$ is a local Hamiltonian of pairs of spins. The
partition function $Z$ is the trace of the density matrix $\rho =
\exp (-\beta H)$ with $\beta=1/T$ the inverse temperature and
$k_B=1$. By means of the Trotter-Suzuki decomposition, the density
matrix can be written as $\rho \simeq [\exp{(-\tau
\sum_{i,j}\hat{H}_{ij})}]^{K+1}$, where $\beta = (K+1) \tau$, and
$\tau$ is the infinitesimal imaginary time slice. Define a local
evolution operator $\hat{U}_{ij} = \exp(-\tau \hat{H}_{ij})$. Then,
the density operator can be represented as $\rho \simeq [\prod_{i,j}
\hat{U}_{ij}]^{K+1} = \prod_{t=1}^{K+1} \prod_{i,j}
\hat{U}^{t}_{ij}$, where $t$ is the Trotter index. In this way, the
density matrix $\rho$ is transformed into a TN. By making a singular
value decomposition (SVD) on $U^{ij}_{i'j'}=\langle
ij|\hat{U}_{ij}|i'j'\rangle $ where $|ij\rangle$ stands for the
direct product basis of spins at site $i$ and $j$, we have
$U^{ij}_{i'j'}=\sum_{s} G^{L}_{ii',s} \lambda^{0}_{s}
G^{R}_{jj',s}$, where $\lambda^{0}$ is the singular value vector,
and $G^{L}$ and $G^{R}$ are two local evolution tensors, each of
which has two physical bonds ($i,i'$ and $j,j'$, respectively) and
one geometrical bond ($s$). For a honeycomb lattice, this step is
depicted in Figs. \ref{fig-TensorNet} (a) and (b). Next, by
contracting the shared bonds among $G^{L}$ and $G^{R}$ [Fig.
\ref{fig-TensorNet} (b)], we get
\begin{eqnarray}
 T^{L}_{il,xyz}=\sum_{jk} G^{L}_{ij,x} G^{L}_{jk,y} G^{L}_{kl,z},
 \ T^{R}_{il,xyz}=\sum_{jk} G^{R}_{ij,x} G^{R}_{jk,y} G^{R}_{kl,z},
\label{eq-InitialT}
\end{eqnarray}
where $x$, $y$ and $z$ are three inequivalent bonds on a honeycomb
lattice [Fig. \ref{fig-TensorNet} (c)]. The density operator $\rho$
at an inverse temperature $\tau$ has the form of
\begin{eqnarray}
 \rho_{\cdots ii'jj' \cdots}=Tr_{G} ( \cdots \lambda^{II}_{y}
 \lambda^{III}_{z} T^{L}_{ii',xyz} \lambda^{I}_{x} T^{R}_{jj',xy'z'}
 \lambda^{II}_{y'} \lambda^{III}_{z'} \cdots),
\label{eq-Initialrho}
\end{eqnarray}
in which $Tr_{G}$ is the trace over all contracted geometrical
bonds, and $\lambda^{I}$, $\lambda^{II}$, $\lambda^{III}$ are three
inequivalent singular value vectors with the initial value
$\lambda^{0}$. This gives a tensor product density operator (TPDO),
which is a direct extension of the matrix product density operator
\cite{MPDO} and the tensor product states. In fact, the TPDO is the
infinite product of two inequivalent tensors $T^{L}$ and $T^{R}$ for
two sublattices (denoted as $\mathcal{SL}_{a}$ and
$\mathcal{SL}_{b}$) of the honeycomb lattice as well as
$\lambda^{I}$, $\lambda^{II}$ and $\lambda^{III}$ for three
inequivalent bonds [Fig. \ref{fig-TensorNet} (c)]. Because of the
structure of the present lattice and the forms of interactions, only
two inequivalent tensors are adequate here, which is independent of
any specific states. For a Kagom\'{e} lattice, at least three
inequivalent tensors are needed. We now present the fLTRG process on
bond $x$ [Fig. \ref{fig-HOOI} (a)] as an example.

\textit{Evolution}.--- By acting $G^{L}$ and $G^{R}$ in pairs on the
TPDO to evolve along the imaginary time direction, we have
\begin{eqnarray}
 \widetilde{T}^{L}_{ik,(xx')yz}=\sum_{j} G^{L}_{ij,x'} T^{L}_{jk,xyz},
 \ \widetilde{T}^{R}_{ik,(xx')yz}=\sum_{j} G^{R}_{ij,x'} T^{R}_{jk,xyz},
\label{eq-Evolve}
\end{eqnarray}
as shown in Fig. \ref{fig-HOOI} (b). The dimension of the newly
gained bond $(xx')$ is enlarged. We denote $(xx')$ as an index
$\alpha$. Then, the corresponding singular value vector
$\widetilde{\lambda}^{I}$ is a direct product of $\lambda^{0}$ and
$\lambda^{I}$: $\widetilde{\lambda}^{I}_{\alpha}=\lambda^{0}_{x}
\lambda^{I}_{x'}$. To obtain the optimal approximation of the TPDO
with the truncation of the enlarged bond, we propose a modified HOOI
(mHOOI) algorithm. The original HOOI takes the interactions among
each mode of a tensor into account by iterating the orthogonal
transformation on every bond, implying that the effect of other
bonds (or modes) is thus considered when truncating one bond. For
our purpose, we suggest the mHOOI algorithm by considering the
interactions of not only each bond but also each tensor.

\textit{mHOOI}.--- Define the reduced matrix $M^{L}$ ($M^{R}$) of
$\widetilde{T}^{L}$ ($\widetilde{T}^{R}$) on bond $\alpha$ by
\begin{eqnarray}
M^{L}_{\alpha \beta}=\sum_{ikyz} \widetilde{T}^{L}_{ik,\alpha yz}
\widetilde{T}^{L}_{ik,\beta yz} \lambda^{II}_{y} \lambda^{III}_{z},
\ M^{R}_{\alpha \beta}=\sum_{ikyz} \widetilde{T}^{R}_{ik,\alpha yz}
\widetilde{T}^{R}_{ik,\beta yz} \lambda^{II}_{y} \lambda^{III}_{z}.
\label{eq-RM}
\end{eqnarray}
Making an eigenvalue decomposition on $M^{L}$ and $M^{R}$, we have
\begin{eqnarray}
M^{L}_{\alpha \beta}=\sum_{\chi} \Lambda^{L}_{\alpha \chi}
\Gamma^{L}_{\chi} \Lambda^{L}_{\beta \chi}, \ M^{R}_{\alpha
\beta}=\sum_{\chi} \Lambda^{R}_{\alpha \chi} \Gamma^{R}_{\chi}
\Lambda^{R}_{\beta \chi}, \label{eq-RMEVD}
\end{eqnarray}
where the matrix $\Lambda^{L}$ ($\Lambda^{R}$) is formed by the
eigenvectors of $M^{L}$ ($M^{R}$), and $\Gamma^{L}$ ($\Gamma^{R}$)
contains the corresponding eigenvalues. The non-orthogonal
transformation matrix $V^{L}$ ($V^{R}$) can be obtained by
\begin{eqnarray}
 V^{L}_{\alpha \chi}=\Lambda^{L}_{\alpha \chi} \sqrt{\Gamma^{L}_{\chi}}, \ \ V^{R}_{\alpha \chi}=\Lambda^{R}_{\alpha \chi} \sqrt{\Gamma^{R}_{\chi}}.
\label{eq-NonOrTM}
\end{eqnarray}
Acting $V^{L}$ and $V^{R}$ to $\widetilde{\lambda}^{I}_{\alpha}$ and
their inverses to $\widetilde{T}^{L}$ and $\widetilde{T}^{R}$,
respectively, as shown in Figs. \ref{fig-HOOI} (c)-(d), one has
\begin{eqnarray}
 && {\cal M}_{\chi \chi'}=\sum_{\alpha} V^{L}_{\alpha \chi} \widetilde{\lambda}^{X}_{\alpha} V^{R}_{\alpha \chi'}, \\
 && \mathcal{T}^{L}_{ik,\chi yz}=\sum_{\alpha} \widetilde{T}^{L}_{ik,\alpha yz} (V^{L})^{-1}_{\alpha
 \chi}, \\
 && \mathcal{T}^{R}_{ik,\chi yz}=\sum_{\alpha} \widetilde{T}^{R}_{ik,\alpha yz} (V^{R})^{-1}_{\alpha
 \chi}.
\label{eq-TrunT}
\end{eqnarray}
It corresponds to inserting two unit matrices [$I=(V^{L})^{-1} \cdot
V^{L}=V^{R} \cdot (V^{R})^{-1} $, as shown in Fig. \ref{fig-HOOI}
(c)] and changes nothing for the TPDO. The intermediate matrix
${\cal M}$ can be decomposed through SVD as
\begin{eqnarray}
 {\cal M}_{\chi \chi'}=\sum_{\kappa} P_{\chi \kappa} \bar{\lambda}_{\kappa} Q_{\chi'
 \kappa},
\label{eq-SVDM}
\end{eqnarray}
where $\bar{\lambda}_{\kappa}$ is the singular value vector arranged
in a descending order, $P$ ($Q$) is formed by the left (right)
singular vectors of ${\cal M}$. Now, we keep the
$D_{c}$ largest singular values as new singular value vector
$\lambda^{I}$ of bond $x$, and normalize $\lambda^{I}$ by dividing
the renormalization factor $r^{I}_{n}=\sqrt{\sum^{D_{c}}_{i=1}
(\lambda^{I}_{i}})^{2}$ with $n$ the step of evolution. Meanwhile,
acting the top $D_{c}$ singular vectors in $P$ and $Q$ on
$\mathcal{T}^{L}$ and $\mathcal{T}^{R}$, respectively, we get new
tensors with truncated bond $x$ [Figs. \ref{fig-HOOI} (e)-(f)]
\begin{eqnarray}
 T^{L}_{ik,\kappa yz}=\sum_{\chi} \mathcal{T}^{L}_{ik,\chi yz} P_{\chi
 \kappa}, \ \
 T^{R}_{ik,\kappa yz}=\sum_{\chi} \mathcal{T}^{R}_{ik,\chi yz} Q_{\chi
 \kappa}.
\label{eq-FinalT}
\end{eqnarray}
Then we renew $\lambda^{II}$, $\lambda^{III}$ and $\lambda^{I}$ in
turn without truncating their dimensions by making the iteration
procedure several times (e.g. five times in our case) according to
the operations described in Figs. \ref{fig-HOOI} (c)-(f) until
reaching a convergence.

\textit{fLTRG step}.--- The evolution and mHOOI processes give a
complete fLTRG step on bond $x$. Doing this step on $x$, $y$ and $z$
bonds in one turn corresponds to that the TPDO is evolved with an
imaginary time $\tau$. After doing the $Kth$-turn, the inverse
temperature for the TPDO reaches $\beta=(K+1)\tau$. Consequently,
the density operator $\rho$ is obtained by Eq.
(\ref{eq-Initialrho}).

It should be remarked that in the above mHOOI procedure, we first
make the truncation on bond $x$ and then do the iteration over three
bonds so that the interactions among bonds and tensors are well
taken into account. Certainly, one may also iterate first and then
truncate the enlarged bond, which gives almost the same result
according to our calculations. However, doing the truncation first
is obviously more efficient. Moreover, for the present case with an
infinite size, we have only three inequivalent bonds on which the
iteration goes. In principle, such an mHOOI may also be applied to
the finite-size systems by sweeping over all inequivalent bonds to
achieve the optimal approximation.

\textit{Free energy}.--- Partition function Z can be obtained by
tracing all physical and geometrical bonds. Tracing all physical
bonds of the TPDO, we get a 2D classical TN (CTN). The free energy
per site $f = - \lim_{N\rightarrow\infty} \ln Z(\beta)/(N\beta)$
with $N$ the number of lattice sites, is comprised of two parts, the
renormalization factors $r_{n}^{\mu}$ and the contributed factor per
site $r_{2d}$ obtained through the contraction of the CTN:
\begin{eqnarray}
 f(\beta)=\frac{1}{2\beta}(\sum^{K}_{n=1}\sum_{\mu=I,II,III}\ln{r^{\mu}_{n}}+ 2\ln
 {r_{2d}}).
\label{eq-FreeE}
\end{eqnarray}
The thermodynamical quantities including energy, magnetization,
susceptibility and specific heat of the 2D quantum systems can thus
be obtained.

What is more, the ground state properties can also be studied with
the fLTRG algorithm. When one takes $K\rightarrow \infty$ and $\tau
\rightarrow 0$, the renormalization factors of each fTRG step
converge to $1$. The ground state energy per site $e_0$ has a
simple form of
\begin{eqnarray}
 e_{0}=\lim_{K\rightarrow \infty}\lim_{\tau\rightarrow0}\frac{1}{2\tau}
 \ln \prod_{\mu=I, II, III}{{r}^{\mu}}.
\label{eq-FreeEg}
\end{eqnarray}

\begin{figure}[tbp]
\includegraphics[angle=0,width=1\linewidth]{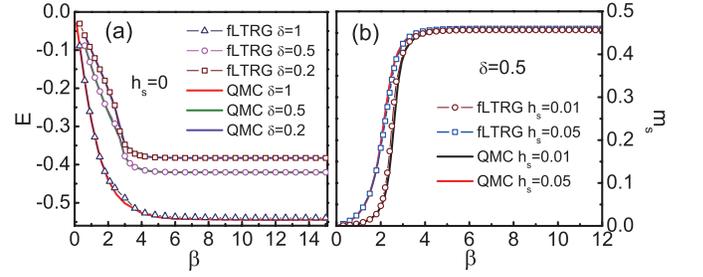}
 \caption{(Color online) The inverse temperature $\beta$ dependence of (a)
 the energy per site $E$ for different $\delta$ with $h_s=0$ and (b) the staggered
 magnetization per site $m_s$ for different $h_s$ with $\delta=0.5$ in
 a spin-$1/2$ anisotropic Heisenberg antiferromagnet on a honeycomb lattice,
 where $D_{c}=22$ and $\tau=0.005$. The QMC results are
 also included for comparison.}
\label{fig-Energy}
\end{figure}

\textit{Spin-1/2 Heisenberg antiferromagnet on honeycomb lattice}.
--- To test the efficiency and accuracy of the fLTRG algorithm, we employ the
spin-$1/2$ anisotropic Heisenberg antiferromagnet on honeycomb
lattice in a staggered magnetic field $h_{s}$, and compare the fLTRG
results with the QMC calculations \cite{QMC}. The local Hamiltonian of nearest-neighbor
spins reads
\begin{eqnarray}
 \hat{H}_{ij}=\delta(\hat{S}^{x}_{i} \hat{S}^{x}_{j}+\hat{S}^{y}_{i} \hat{S}^{y}_{j})
 +\hat{S}^{z}_{i} \hat{S}^{z}_{j}+(\hat{S}^{z}_{i}-\hat{S}^{z}_{j})h_{s}/3,
\label{eq-Hamiltonian}
\end{eqnarray}
where $\hat{S}_{i}^{x}$, $\hat{S}_{i}^{y}$ and $\hat{S}_{i}^{z}$ are the $x$-, $y$-
and $z$-component of spin operator on the $i$th site, respectively,
and $\delta$ measures the anisotropy of spin couplings. The energy
per site can be calculated by $E=-d(\beta f)/d\beta$, and the
staggered magnetization per site is obtained by
$m_{s}=\partial{f}/\partial{h_{s}}$. Fig. \ref{fig-Energy} gives $E$
and $m_s$ as functions of the inverse temperature $\beta$ for
different $\delta$ with $h_s=0$ and different $h_s$ with
$\delta=0.5$, respectively. It can be seen that our fLTRG results
are in nice agreement with those of QMC calculations, showing
that the fLTRG algorithm is feasible, efficient and accurate.

\begin{figure}[tbp]
\includegraphics[angle=0,width=0.8\linewidth]{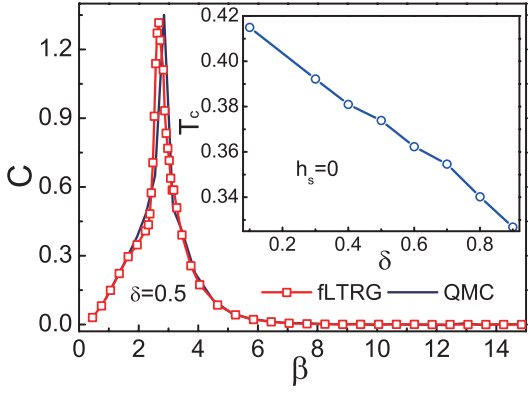}
 \caption{(Color online) The inverse temperature $\beta$ dependence of the
 specific heat for $\delta=0.5$ and $h_s=0$ in
 a spin-$1/2$ anisotropic Heisenberg antiferromagnet on a honeycomb lattice.
 The QMC result is included for
 a comparison. Inset gives the $\delta$ dependence of the critical
 temperature $T_c$. Here $D_c=32$.}
\label{fig-Phase}
\end{figure}

The specific heat as a function of $\beta$ is calculated by
$C=-\beta^{2}dE/d\beta$, as shown in Fig. \ref{fig-Phase} for
$\delta=0.5$. A divergent peak at a critical temperature $T_c$ is
observed, which indicates that a phase transition occurs between a
paramagnetic phase and an antiferromagnetic phase at $T_c$. It is
also well consistent with the QMC result, showing again the
efficiency and accuracy of the fLTRG method. In the inset of Fig.
\ref{fig-Phase}, $T_c$ as a function of $\delta$ is given,
indicating that $T_c$ declines almost linearly with increasing
$\delta$. It should be pointed out that as $\delta \rightarrow 1$,
the divergent peak of the specific heat becomes gradually round
owing to the increase of quantum fluctuations, and the phase
transition no longer exists at $\delta=1$, being consistent with the
Mermin-Wagner theorem.

\begin{figure}[tbp]
\includegraphics[angle=0,width=1\linewidth]{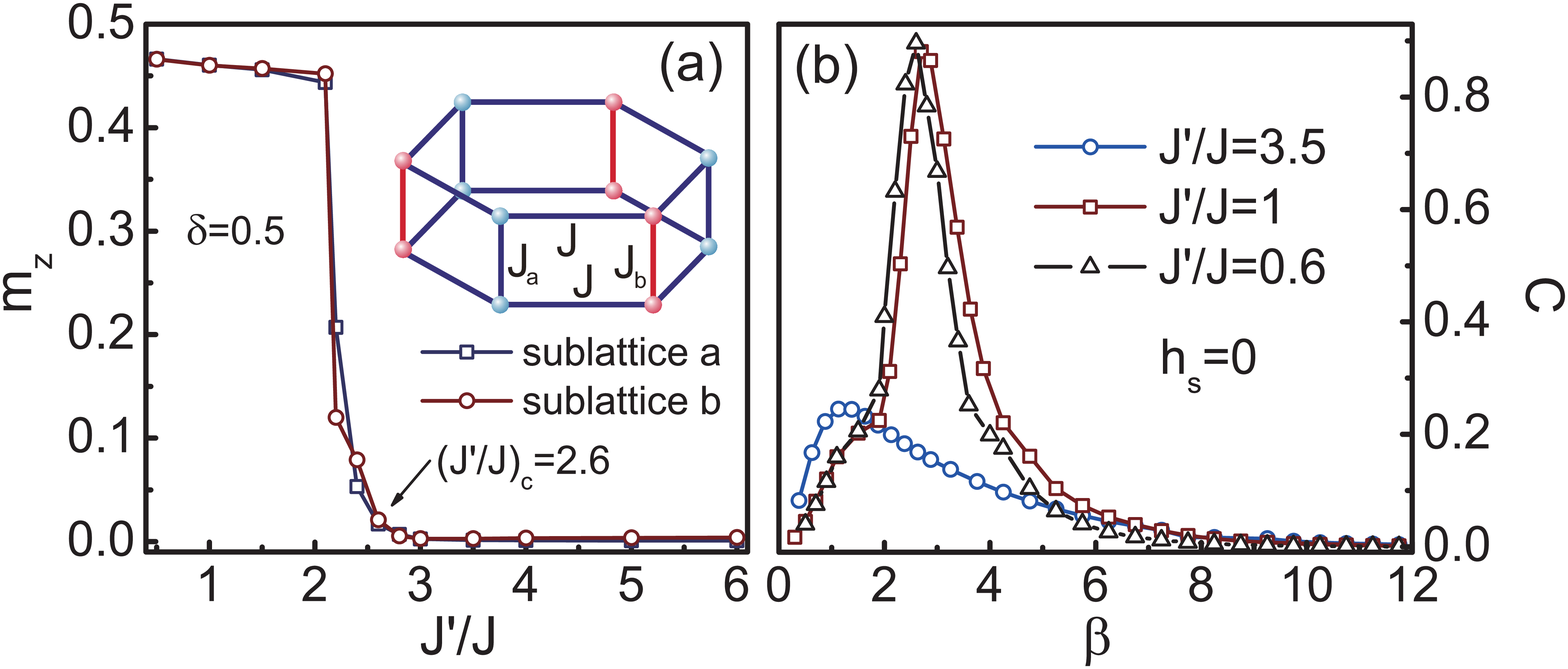}
 \caption{(Color online) (a) The sublattice magnetization per site
 $m_z$ as a function of $J'/J$ for a spin-$1/2$ frustrated bilayer
 honeycomb anisotropic Heisenberg model (inset) with $\delta=0.5$ at
 zero temperature, where a quantum critical point $(J'/J)_{c}=2.60(2)$
 is identified. (b) The inverse temperature $\beta$ dependence of
 the specific heat for different
 $J'/J$ with $h_s=0$. Here $D_{c}=32$.}
\label{fig-BiMz}
\end{figure}

\textit{Spin-1/2 frustrated bilayer honeycomb Heisenberg model}.
--- To show the power of the fLTRG algorithm, we now apply it to
investigate the spin-$1/2$ anisotropic Heisenberg antiferromagnetic
model on a bilayer honeycomb lattice with alternating
antiferromagnetic and ferromagnetic interlayer interactions $J_a$
and $J_b$ [the inset of Fig. \ref{fig-BiMz} (a)]. It is a quasi-2D
quantum frustrated spin system to which the QMC is hardly accessible
owing to the negative sign problem. The local Hamiltonian of this
model is defined as
\begin{eqnarray}
 \hat{H}_{ij}=\hat{H}^{(1)}_{ij}+\hat{H}^{(2)}_{ij}+(\hat{H}^{(a)}_{i}
 +\hat{H}^{(b)}_{j})/3,
\label{eq-BiHamiltonian}
\end{eqnarray}
where
$\hat{H}^{(\gamma)}_{ij}=J_{\gamma}[\delta_{\gamma}(\hat{S}^{x}_{i}
\hat{S}^{x}_{j}+\hat{S}^{y}_{i} \hat{S}^{y}_{j})+\hat{S}^{z}_{i}
\hat{S}^{z}_{j}]$, with the layer index $\gamma=1$ and $2$, and
$i,j$ the nearest neighbor sites within the single layer;
$\hat{H}^{(a,b)}_{i}=J_{a,b}[\delta_{a,b}(\hat{S}^{x}_{i}
\hat{S}^{x}_{i}+\hat{S}^{y}_{i} \hat{S}^{y}_{i})+\hat{S}^{z}_{i}
\hat{S}^{z}_{i}]$ is the interlayer couplings. When $J_a>0$ and
$J_b<0$, it gives rise to the spin frustration. Without losing
generality, we shall take $J_{1}=J_{2}=J>0$, $J_a>0$, $J_b<0$,
$J_a=-J_b=J'$, and $\delta_{1,2}=\delta_{a,b}=\delta=0.5$. As the
frustration exists, this model would be expected in proper
circumstances to have a QSL ground state that is currently under an
active debate \cite{Lee,SL,SL1}.

Figure \ref{fig-BiMz} (a) shows the sublattice magnetization per
site $m_z$ as a function of the coupling ration $J'/J$ at zero
temperature. It can be seen that there exists a quantum critical
point $(J'/J)_{c}=2.6$, at which a QPT occurs. When $J'/J<2.6$,
$m_z$ is nonzero and decreases slowly with the increase of $J'/J$,
showing that in this regime the system is in an antiferromagnetic
ordered state; around $J'/J \simeq 2.6$, $m_z$ drops sharply; and it
goes to zero for $J'/J>2.6$, while the magnetization in the $x-y$
plane is found about $10^{-3}$ in the whole region, suggesting that
the system enters into a disordered state. This disordered state is
nothing but a gapless QSL state (see below). The reason is that, for
$J'/J>2.6$, the frustration becomes stronger \cite{note}, which
strongly suppresses the magnetic long-range ordering, giving rise to
a QSL state.
\begin{figure}[tbp]
\includegraphics[angle=0,width=1\linewidth]{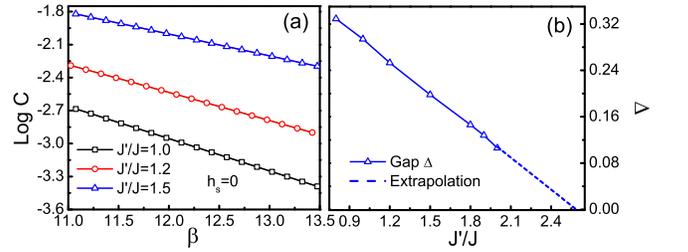}
 \caption{(Color online) (a) Log $C$ versus $\beta$ of the spin-1/2
 frustrated bilayer honeycomb Heisenberg model for
 various $J'/J$ with $h_s=0$; (b) The gap $\Delta$ versus
 $J'/J$, where the extrapolation shows that
 at $(J'/J)_{c}=2.60(2)$, the gap vanishes, indicating a QPT to a gapless QSL state.
 Here $D_{c}=35$.}
\label{fig-LogC}
\end{figure}

The temperature dependence of the specific heat ($C$) of this
frustrated spin system is given for different $J'/J$ in Fig.
\ref{fig-BiMz} (b). When $J'/J<(J'/J)_{c}=2.6$, the specific heat
displays a divergent peak at a critical temperature for a given
$J'/J$, showing a second-order phase transition between a
paramagnetic state and an Ising-type ordered state. It appears that
the critical temperature depends weakly on $J'/J$. For
$J'/J>(J'/J)_{c}$, the specific heat shows a round peak, and no
phase transition happens, which is consistent with the observation
that in this regime the ground state of system is in a gapless QSL
state.

Finally, we observe that, when $J'/J<2.6$, this frustrated bilayer
spin model is in an Ising-type ordered state with a gap. It is
evidenced by the low-temperature behavior of the specific heat that
decays exponentially with the inverse temperature $\beta$ in a form
of $C \thicksim \exp{(-\Delta\beta)}$. The gap $\Delta$ can be
determined by using a linear fitting between $\log{C}$ and $\beta$
for different $J'/J$ with $h_s=0$, as presented in Fig.
\ref{fig-LogC} (a), showing a perfect linear $J'/J$ dependence of
the gap [Fig. \ref{fig-LogC} (b)]. By extrapolation, one may observe
that $\Delta$ vanishes at $(J'/J)_{c}=2.60(2)$, confirming the QPT
from a gapped Ising-type ordered state to a gapless QSL state when
the frustration effect becomes stronger.

In conclusion, by extending the Tucker decomposition to a tensor
network, we propose a novel algorithm coin as the fLTRG, and examine
its efficiency and accuracy by employing a spin-$1/2$ Heisenberg
antiferromagnet on honeycomb lattice. The fLTRG results are well in
agreement with the QMC calculations. To show the power of the fLTRG
algorithm, it is applied to a spin-$1/2$ frustrated bilayer
honeycomb Heisenberg model with alternating interlayer couplings,
and a quantum phase transition is disclosed, where a gapless quantum
spin liquid phase is identified. The present fLTRG algorithm could
be straightforwardly extended to other quantum lattice systems.

We are indebted to F. Yei, Q. R. Zheng, X. Yan, B. Xi, Y. Zhao and
Z. Zhang for stimulating discussions. This work is supported in part
by the NSFC (Grant Nos. 90922033 and 10934008), the MOST of China
(Grant No. 2012CB932901) and the CAS.

\end{document}